\documentstyle[12pt,epsf]{article}
\def\beq{\begin{equation}}
\def\eeq{\end{equation}}
\def\bea{\begin{eqnarray}}
\def\eea{\end{eqnarray}}
\def\bq{\begin{quote}}
\def\eq{\end{quote}}

\def\bq{\begin{quote}}
\def\eq{\end{quote}}

\def\bq{\begin{quote}}
\def\eq{\end{quote}}

\def\gappeq{\mathrel{\rlap {\raise.5ex\hbox{$>$}}
{\lower.5ex\hbox{$\sim$}}}}

\def\lappeq{\mathrel{\rlap{\raise.5ex\hbox{$<$}}
{\lower.5ex\hbox{$\sim$}}}}

\def\bbz{fa Z \kern-8.9pt Z}

\begin{document}

\baselineskip 24pt
\newcommand{\sheptitle}
{Inverted Hierarchy Models of Neutrino Masses}

\newcommand{\shepauthor}
{S. F. King and N. Nimai Singh
\footnote{On leave from the Department of Physics,
Gauhati University, Guwahati - 781014, India.}}

\newcommand{\shepaddress}
{Department of Physics and Astronomy,
University of Southampton, Southampton, SO17 1BJ, U.K.}

\newcommand{\shepabstract}
{We study models of neutrino masses which naturally give rise to
an inverted mass hierarchy and bi-maximal mixing. The models are 
based on the see-saw
mechanism with three right-handed neutrinos, which 
generates a single mass term
of the form $\nu_e(\nu_{\mu}+\nu_{\tau})$ corresponding
to two degenerate neutrinos $\nu_e$ and $\nu_{\mu}+\nu_{\tau}$,
and one massless neutrino $\nu_{\mu}-\nu_{\tau}$.
Atmospheric neutrino oscillations are accounted for if the
degenerate mass term is about $5\times 10^{-2}$ eV.
Solar neutrino oscillations of the Large Mixing Angle MSW type
arise when small perturbations are included leading
to a mass splitting between the degenerate pair of 
about $(1.7-2.0)\times 10^{-4}$ eV for the successful cases.
We study the conditions that such models must satisfy in
the framework of a $U(1)$ family symmetry broken by vector singlets,
and catalogue the simplest examples. We then perform a
renormalisation group analysis of the neutrino masses and mixing
angles, assuming the supersymmetric standard model,
and find modest radiative corrections of a few per cent,
showing that the model is stable.
At low energies we find $\sin^22\theta_{23}\approx 0.93-0.96$
and $\sin^22\theta_{12}\approx 0.9-1.0$.}

\begin{titlepage}
\begin{flushright}
hep-ph/0007243\\
\end{flushright}
\begin{center}
{\large{\bf \sheptitle}}
\bigskip \\ \shepauthor \\ \mbox{} \\ {\it \shepaddress} \\ \vspace{.5in}
{\bf Abstract} \bigskip \end{center} \setcounter{page}{0}
\shepabstract
\end{titlepage}

\section{Introduction}
The latest atmospheric neutrino results based on 1117 days of data
from Super Kamiokande are still consistent with a standard two neutrino
oscillation $\nu_{\mu}\rightarrow \nu_{\tau}$
with a near maximal mixing angle $\sin^2 2\theta_{23} >0.88$ and
a mass square splitting $\Delta m_{23}^2$ from $1.5\times 10^{-3}$ to $5\times 10^{-3}\ eV^2$
at 90\% CL \cite{sobel}. The sterile neutrino oscillation
hypothesis $\nu_{\mu}\rightarrow \nu_{s}$ is
excluded at 99\% CL.

Super Kamiokande is also beginning to provide important clues
concerning the correct solution to the solar neutrino problem. The
latest results from 1117 days of data from Super Kamiokande \cite{suzuki}
sees a one sigma day-night asymmetry, and a flat energy spectrum,
which together disfavour the small mixing angle (SMA) MSW solution
\cite{MSW},
the just-so vacuum oscillation hypothesis \cite{VO}
and the sterile neutrino hypotheses. 
All three possibilities are now excluded at
95\% CL. The results allow much of the large mixing angle (LMA) MSW
region \cite{MSW}, which now looks like the leading candidate for the
solution to the solar neutrino problem. For example a typical
point in the LMA MSW region is
$\sin^22\theta_{12} \approx 0.75$ and $\Delta m_{12}^2 \approx 2.5\times
10^{-5}\ eV^2$\cite{LMAMSW}.

\begin{figure}[t]
\epsfxsize=10cm
\epsfxsize=10cm
\hfil
\epsfbox{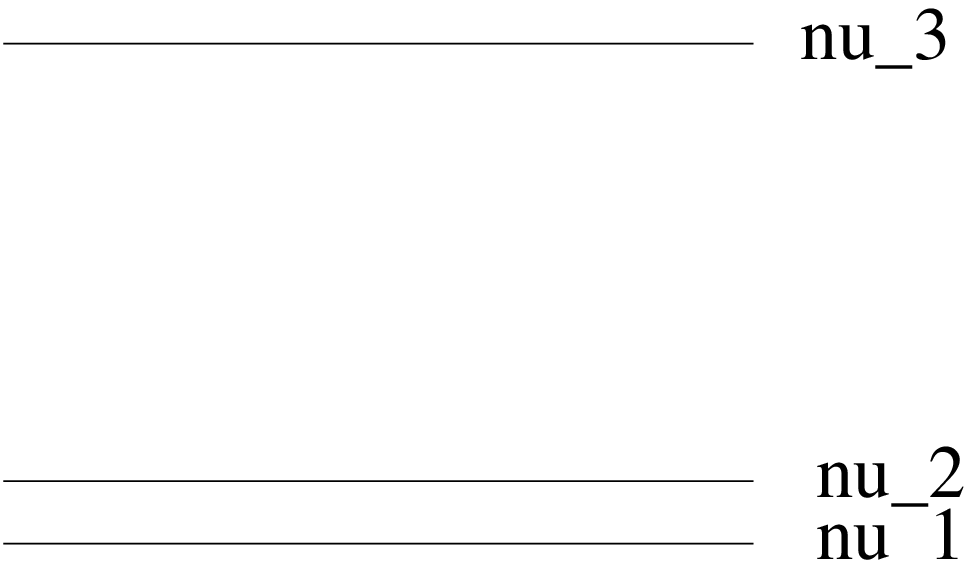}
\hfil
\caption{{\footnotesize A
conventional neutrino mass hierarchy
$m_{\nu_1}\ll m_{\nu_2} \ll m_{\nu_3}$.}} \label{hierarchy}
\end{figure}

\begin{figure}[t] 
\epsfxsize=10cm 
\epsfxsize=10cm
\hfil 
\epsfbox{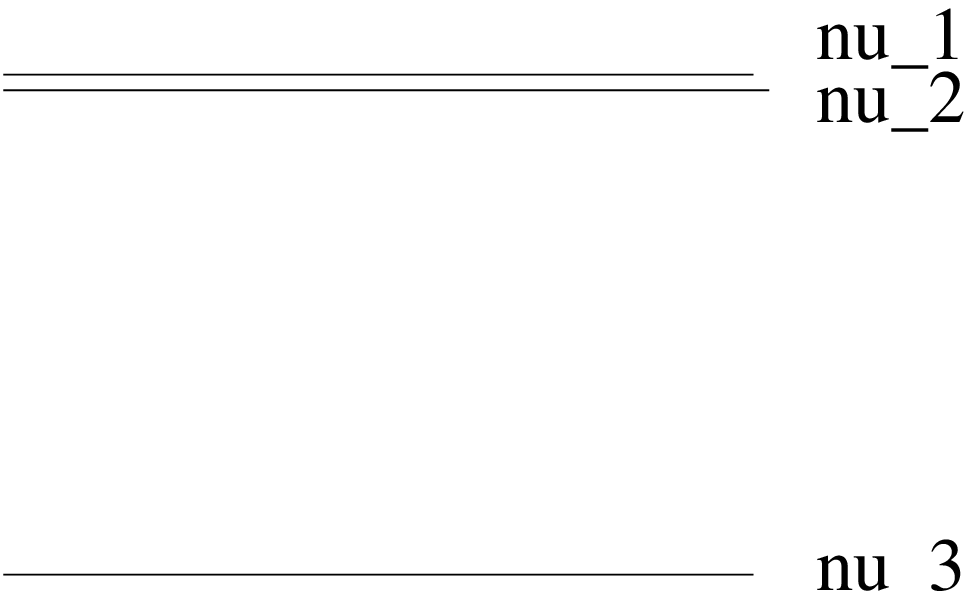} 
\hfil 
\caption{{\footnotesize An inverted neutrino mass hierarchy
$m_{\nu_3}\ll m_{\nu_2} \approx m_{\nu_1}$.
Note that the mass splitting $|m_{\nu_1}-m_{\nu_2}|$ must be
much smaller than in the hierarchical case (as discussed later in the
text.)}}
\label{invertedhierarchy}
\end{figure}

Once the CHOOZ constraint \cite{CHOOZ} is taken into account,
the latest results seem to imply an approximate bi-maximal mixing scenario
$\tan \theta_{23}\approx 1$, $\tan \theta_{12}\approx 1$, $\theta_{13}\ll 1$
relating
the three standard neutrinos $\nu_e, \nu_{\mu}, \nu_{\tau}$,
to the three mass eigenvalues $m_{\nu_1},m_{\nu_2},m_{\nu_3}$,
where only the values of $|\Delta m_{23}^2|=|m_{\nu_3}^2-m_{\nu_2}^2|$ and
$|\Delta m_{12}^2|=|m_{\nu_2}^2-m_{\nu_1}^2|$
are determined by current experiments.
We do not know if the
neutrino masses are hierarchical or approximately degenerate
with small mass splittings. Another possibility
is that the neutrinos have an inverted mass hierarchy,
i.e. two of them are approximately degenerate with a small mass
splitting, while the third is much lighter.
The conventional and inverted neutrino mass hiearchies are
depicted schematically in Figures \ref{hierarchy} and 
\ref{invertedhierarchy}. 
Neutrino factories will be able to determine the sign of
$\Delta m_{23}^2$ and hence distinguish between the two cases
\cite{golden}.

The see-saw mechanism \cite{seesaw} implies that the three light
neutrino masses arise from some large mass scales corresponding to
the Majorana masses of some heavy ``right-handed neutrinos''
$N^p_R$ $M^{pq}_{RR}$ ($p,q=1,\cdots ,Z$) whose entries take
values which extend from say $\sim 10^{16}$ GeV down to perhaps
several orders of magnitude lower. The presence of electroweak
scale Dirac mass terms $m_{LR}^{ip}$ (a $3 \times Z$ matrix)
connecting the left-handed neutrinos $\nu^i_L$ ($i=1,\ldots 3$) to
the right-handed neutrinos $N^p_R$ then results in a very light
see-saw suppressed effective $3\times 3$ Majorana mass matrix
\beq
m_{LL}=m_{LR}M_{RR}^{-1}m_{LR}^T \label{seesaw}
\eeq
for the
left-handed neutrinos $\nu_L^i$, which are the light physical
degrees of freedom observed by experiment. If the neutrino masses
arise from the see-saw mechanism then it is natural to assume the
existence of a physical neutrino mass hierarchy 
$m_{\nu_1}\ll m_{\nu_2} \ll m_{\nu_3}$, 
which implies $\Delta m_{23}^2 \approx
m_{\nu_3}^2$, and $\Delta m_{12}^2 \approx m_{\nu_2}^2$, which
fixes $m_{\nu_3} \approx 5 \times 10^{-2}eV$, and (assuming the
LMA MSW solution) $m_{\nu_2} \approx 5 \times 10^{-3}eV$, with
rather large errors. Thus $m_{\nu_2}/m_{\nu_3} \sim 0.1$.

Hierarchical neutrino masses are not guaranteed because
a hierarchy in $m_{LR}$ does not necessarily imply a hierarchy in
$m_{LL}$ since $M_{RR}$ is also expected to be hierarchical,
and the hierarchies may cancel out in the see-saw mechanism.
Nevertheless it would seem surprising if the resulting neutrino
spectrum came out to be degenerate
$m_{\nu_1}\approx m_{\nu_2} \approx m_{\nu_3}$
with three accurately equal masses
of say an eV but with small mass splittings appropriate to the
atmospheric and solar data especially taking into account
radiative corrections \cite{Ellis}, although such a scenario could be enforced
by a symmetry \cite{Ross}. The possibility of an
inverted mass hierarchy $m_{\nu_3}\ll m_{\nu_2} \approx m_{\nu_1}$
with bi-maximal mixing however is more natural, since it
follows from a simple form of mass matrix \cite{inverted1}
\beq
m_{LL}\sim
\left( \begin{array}{ccc}
0 & b & c\\
b & 0 & 0\\
c & 0 & 0
\end{array}
\right)
\label{inverted}
\eeq
which corresponds to neutrino masses of the form
$\nu_e(b\nu_{\mu}+c\nu_{\tau})$, which implies
two degenerate neutrinos $\nu_e$ and $b\nu_{\mu}+c\nu_{\tau}$,
and a massless neutrino $c\nu_{\mu}-b\nu_{\tau}$,
with neutrino mixing angles
$\tan \theta_{23}=c/b$, $\tan \theta_{12}=1$, $\theta_{13}=0$.
It has also been pointed out that the mass matrix in
Eq.\ref{inverted} can be generated from the see-saw mechanism
using two right-handed neutrinos with an off-diagonal
Majorana mass, using a chiral $U(1)$ family
symmetry to give the desired textures \cite{inverted2}.

Our discussion of the inverted hiearchy models differs
from that already presented in the literature in 
the following three ways. Firstly we consider the see-saw mechanism in
the framework of the supersymmetric standard model with {\em three}
right-handed neutrinos, and focus on a particular texture for
the heavy right-handed neutrinos 
which is capable of approximately reproducing
the mass matrix in Eq.\ref{inverted}.
Secondly we introduce
a {\em vector} $U(1)$ family symmetry which is broken by vector-like
singlets, which allows both signs of charges to contribute,
and make a computer scan over all charges which
can lead to the desired mass matrix, and tabulate the simplest
cases. Thirdly we perform a renormalisation group (RG) analysis
of some cases, in order to examine the radiative
corrections in going from the GUT scale to low energy.

In section 2 we give an analytic discussion of the model,
and in section 3 we introduce a $U(1)$ family symmetry and tabulate
the simplest charges consistent with our scheme.
Section 4 contains a renormalisation group analysis of some
examples, and section 4 concludes the paper.

\section{Analytic Discussion}
We begin our discussion by returning to the case of
a conventional neutrino mass hierarchy where
the presence of a large 23 mixing angle looks a bit surprising
at first sight, especially given our experience with small
quark mixing angles. Several explanations have been proposed
\cite{several}, but the simplest idea is that
the contributions to the 23 block of the light effective Majorana matrix
come predominantly from a single right-handed neutrino,
which causes the 23 subdeterminant to approximately
vanish. A dominant single right-handed neutrino then naturally leads
to a hierarchical neutrino mass spectrum, with a single dominant physical
neutrino mass and two much lighter neutrinos. This mechanism, called
single right-handed neutrino dominance (SRHND), was
proposed in \cite{SK}, and developed for bi-maximal mixing in
\cite{SK2}. The effect of radiative corrections on the bi-maximal
SRHND case was considered in \cite{KS}.
We then show that for the off-diagonal heavy Majorana
texture, a reversal of the SRHND conditions leads to an inverted mass
hierarchy and bi-maximal mixing.

We first write the neutrino Yukawa matrix in general (in the LR basis) as
\beq
Y_{\nu}=
\left( \begin{array}{ccc}
a' & a & d\\
b' & b & e\\
c' & c & f
\end{array}
\right)
\label{N3Dirac}
\eeq
There are now three distinct textures for the heavy Majorana neutrino
matrix which maintain the isolation of the dominant right-handed
neutrino $N_{R3}$, namely the diagonal,
democratic and off-diagonal textures
introduced previously\cite{SK,SK2}. Only one of these textures,
namely the off-diagonal case, can lead to an inverted mass hierarchy.
We therefore specialise to the off-diagonal heavy Majorana texture:
\footnote{Actually this texture represents a set of off-diagonal
Majorana textures related by a re-ordering of the right-handed
neutrino fields. For example if we cyclically permute 
the right-handed neutrino
fields as $1\rightarrow 2,2\rightarrow 3,3\rightarrow 1$
then we would have
$$
Y_{\nu}=
\left( \begin{array}{ccc}
d & a' & a \\
e & b' & b \\
f & c' & c 
\end{array}
\right), \ \ M_{RR}=
\left( \begin{array}{ccc}
Y & 0 & 0    \\
0 & 0 & X \\
0 & X & 0
\end{array}
\right)
$$
with $m_{LL}$ unchanged. The results in this paper therefore
apply to this larger class of model, corresponding to all possible
re-orderings of right-handed neutrino fields.
For example the matrices $Y_{\nu}$ and $M_{RR}$ in 
Tables 3,4 may be permuted so that the large elements
of order unity appear in the 23 and 33 positions of $Y_{\nu}$,
which may be more natural from the perspective
of unified theories where a large 33 element is expected.}
\beq
M_{RR}=
\left( \begin{array}{ccc}
0 & X & 0    \\
X & 0 & 0 \\
0 & 0 & Y
\end{array}
\right)
\label{MRRoff-diag}
\eeq
From Eqs.\ref{N3Dirac} and \ref{MRRoff-diag},
the see-saw formula Eq.\ref{seesaw} implies
\beq
m_{LL}
 =
\left( \begin{array}{ccc}
\frac{d^2}{Y}+\frac{2aa'}{X}
& \frac{de}{Y} +\frac{a'b}{X}+\frac{ab'}{X}
& \frac{df}{Y}+\frac{a'c}{X}+\frac{ac'}{X}    \\
.
& \frac{e^2}{Y} +\frac{2bb'}{X}
& \frac{ef}{Y} +\frac{b'c}{X}+\frac{bc'}{X}   \\
.
& .
& \frac{f^2}{Y} +\frac{2cc'}{X}
\end{array}
\right){v_2^2}
\label{matrix5}
\eeq
where $v_2$ is the Higgs vacuum expection value which relates the
neutrino mass matrix to the neutrino Yukawa matrix
$m_{LR}=Y_{\nu}v_2$.
From this starting point we may obtain either a conventional
hierarchy or an inverted hiearchy, depending on the conditions
applied to the couplings.

\begin{itemize}
\item 
The conditions for a conventional hierarchy 
as in Fig.\ref{hierarchy} via SRHND are
simply that the third right-handed neutrino dominates
the 23 block of $m_{LL}$,
\beq
\frac{e^2}{Y} \sim \frac{ef}{Y} \sim \frac{f^2}{Y} \gg \frac{xx'}{X}
\label{SRHNDoffdiag}
\eeq
where $x \in a,b,c$ and $x' \in a',b',c'$.
For example in the limit that only $e\approx f$ are non-zero
the spectrum consists of a decoupled massless $\nu_e$,
plus another massless state $\nu_{\mu}-\nu_{\tau}$
(because the 23 subdeterminant
vanishes) giving maximal atmospheric mixing with the massive
state $\nu_{\mu}+\nu_{\tau}$. The milder conditions in Eq.\ref{SRHNDoffdiag}
allow the massless degeneracy to be broken, and 
$\nu_e \rightarrow \nu_{\mu}-\nu_{\tau}$
solar oscillations.
Assuming SRHND 
the contribution to the lepton 23 and 13
mixing angles from the neutrino sector are approximately
\beq
\tan \theta_{23} \approx \frac{e}{f}, \ \
\tan \theta_{13} \approx \frac{d}{\sqrt{e^2+f^2}}, \ \
\eeq
so that Super-Kamiokande and CHOOZ \cite{CHOOZ} imply
\beq
d\ll e \approx f
\label{1st}
\eeq
The 12 mixing angle is controlled by the sub-dominant
right-handed neutrinos, and the condition for a $\tan \theta_{12}\sim 1$ is
\cite{SK2}:
\beq
max\left(\frac{a'b}{X},\frac{ab'}{X},\frac{a'c}{X},\frac{ac'}{X}\right)
\sim
max\left(\frac{bb'}{X},\frac{b'c}{X},\frac{bc'}{X},\frac{cc'}{X}\right)
\label{LMAMSWoffdiag}
\eeq

\item
In order to achieve an inverted hiearchy 
as in Fig.\ref{invertedhierarchy} from the off-diagonal texture
in Eq.\ref{MRRoff-diag}, we must require contributions
to $m_{LL}$ in Eq.\ref{matrix5}
such that Eq.\ref{inverted} is approximately reproduced.
It is straightforward to show that only an off-diagonal
texture such as Eq.\ref{MRRoff-diag} allows this. 
This immediately implies that the first and second right-handed
neutrinos with mass term $X$ must give the 
dominant contribution to the matrix relative to the contributions
from the third right-handed neutrino with mass $Y$,
\beq
\frac{(e+d+f)^2}{Y}\ll max\left(\frac{(a'+b'+c')(a+b+c)}{X}\right)
\label{invertedcondn}
\eeq
which is the opposite of the SRHND condition Eq.\ref{SRHNDoffdiag}.
Furthermore we require one of the following conditions to be satisfied
\beq
a',b,c \gg a,b',c', \ \ or \ \  a',b,c \ll a,b',c'. 
\label{invertedcondn2}
\eeq
\end{itemize}

The simplest example which generates an inverted hierarchy is to take
the limit that only $a',b,c$ are non-zero, so that $m_{LL}$ becomes
\beq
m_{LL}\sim
\left( \begin{array}{ccc}
0 & b & c\\
b & 0 & 0\\
c & 0 & 0
\end{array}
\right)\frac{a'v_2^2}{X}
\label{inverted2}
\eeq
which is of the form in Eq.\ref{inverted}.
In order to split the degeneracy, we must allow for small
perturbations. Suppose to begin with
that $d=e=f=0$, then only two right-handed
neutrinos contribute and 
\beq
m_{LL}
 =
\left( \begin{array}{ccc}
\frac{2aa'}{X}
& \frac{a'b}{X}+\frac{ab'}{X}
& \frac{a'c}{X}+\frac{ac'}{X}    \\
.
& \frac{2bb'}{X}
& \frac{b'c}{X}+\frac{bc'}{X}   \\
.
& .
& \frac{2cc'}{X}
\end{array}
\right){v_2^2}
\label{matrix6}
\eeq
Allowing $a,b',c'$ to be non-zero but maintaining
$a',b,c \gg a,b',c'$ we find neutrino masses
\footnote{In fact the eigenvalues have the form
$(|m_{\nu_1}|,-|m_{\nu_2}|,|m_{\nu_3}|)$, with 
$|m_{\nu_1}|\approx |m_{\nu_2}| \gg |m_{\nu_3}|$,
but we can always
redefine the masses to be positive, and this is assumed in
Eq.\ref{14}, and subsequently.
Note that we take the convention $|m_{\nu_1}|>|m_{\nu_2}|$,
which fixes the ordering of the 1st two columns of $V_{MNS}$.}
\beq
m_{\nu_3}=0, 
m_{\nu_2} \approx m_{\nu_1}\approx \frac{a'\sqrt{b^2+c^2}v_2^2}{X}, 
\ \  
(m_{\nu_1}-m_{\nu_2})\approx \frac{2(a'a+b'b+c'c)v_2^2}{X} 
\label{14}
\eeq
and mixing angles
\beq
\tan \theta_{23} \approx \frac{c}{b}, \ \
\theta_{13} \approx \frac{c'b-b'c}{a'\sqrt{b^2+c^2}}, \ \
\tan \theta_{12}\approx 1
\eeq
Note that in the inverted hierarchy case
$|\Delta m_{23}^2| \approx m_{\nu_1}^2 \approx m_{\nu_2}^2$, 
which fixes $m_{\nu_1} \approx m_{\nu_2} \approx 
5 \times 10^{-2}eV$. Since 
$|\Delta m_{12}^2| \approx m_{\nu_1}^2-m_{\nu_2}^2$, 
the LMA MSW solution implies
$|m_{\nu_1}- m_{\nu_2}| \approx 2.5 \times 10^{-4}eV$, with
rather large errors, which is much smaller than the corresponding
mass splitting $5 \times 10^{-3}eV$ in the conventional hierarchy case.
This implies that the perturbations in the inverted hierarchy case
must be much smaller than in the conventional hierarchy case.
A convenient way to describe such perturbations is in the framework
of $U(1)$ family symmetry to which we now turn.

\section{$U(1)$ Family Symmetry}
Introducing a $U(1)$ family symmetry \cite{FN}, \cite{textures},
\cite{IR}, \cite{Ramond} provides a convenient way to
organise the hierarchies within the various Yukawa matrices.
For definiteness we shall focus on a particular class of model based
on a single pseudo-anomalous $U(1)$ gauged family symmetry \cite{IR}.
We assume that the $U(1)$ is broken by the equal VEVs of two
singlets $\theta , \bar{\theta}$ which have vector-like
charges $\pm 1$ \cite{IR}.
The $U(1)$ breaking scale is set by $<\theta >=<\bar{\theta} >$
where the VEVs arise from a Green-Schwartz mechanism \cite{GS}
with computable Fayet-Illiopoulos $D$-term which
determines these VEVs to be one or two orders of magnitude
below $M_U$. Additional exotic vector matter with
mass $M_V$ allows the Wolfenstein parameter \cite{Wolf}
to be generated by the ratio \cite{IR}
\beq
\frac{<\theta >}{M_V}=\frac{<\bar{\theta} >}{M_V}= \lambda \approx 0.22
\label{expansion}
\eeq

The idea is that at tree-level the $U(1)$ family symmetry
only permits third family Yukawa couplings (e.g. the top quark
Yukawa coupling). Smaller Yukawa couplings are generated effectively
from higher dimension non-renormalisable operators corresponding
to insertions of $\theta$ and $\bar{\theta}$ fields and hence
to powers of the expansion parameter in Eq.\ref{expansion},
which we have identified with the Wolfenstein parameter.
The number of powers of the expansion parameter is controlled
by the $U(1)$ charge of the particular operator.
The fields relevant to neutrino masses are the
lepton doublets $L_i$, the charge conjugated right-handed neutrinos
and charged leptons $N^c_i$, $E^c_i$, 
up-type Higgs doublet $H_u$, and a singlet Higgs field
whose VEV signals the heavy Majorana masses $\Sigma$.
These are assigned $U(1)$ charges $l_i$, $n_i$, $e_i$, $h_u=0$,
$\sigma$, respectively. From Eq.\ref{expansion},
the neutrino Yukawa couplings and Majorana mass
terms may then be expanded in powers of the Wolfenstein parameter,
\beq
M_{RR} \sim
\left( \begin{array}{ccc}
\lambda^{|2n_1+\sigma|} & \lambda^{|n_1+n_2+\sigma|}
& \lambda^{|n_1+n_3+\sigma|}\\
. & \lambda^{|2n_2+\sigma|} & \lambda^{|n_2+n_3+\sigma|} \\
.  & .  & \lambda^{|2n_3+\sigma|}
\end{array}
\right) <\Sigma >
\label{mRR}
\eeq
The neutrino Yukawa matrix is explicitly
\beq
Y_{\nu} \sim
\left( \begin{array}{ccc}
\lambda^{|l_1+n_1|} & \lambda^{|l_1+n_2|}
& \lambda^{|l_1+n_3|}\\
\lambda^{|l_2+n_1|} & \lambda^{|l_2+n_2|}
& \lambda^{|l_2+n_3|}\\
\lambda^{|l_3+n_1|} & \lambda^{|l_3+n_2|}
& \lambda^{|l_3+n_3|}
\end{array}
\right)
\label{Ynu}
\eeq
which may be compared to the notation in Eq.\ref{N3Dirac}.
The charged lepton Yukawa matrix is given by
\beq
Y_{e} \sim
\left( \begin{array}{ccc}
\lambda^{|l_1+e_1|} & \lambda^{|l_1+e_2|}
& \lambda^{|l_1+e_3|}\\
\lambda^{|l_2+e_1|} & \lambda^{|l_2+e_2|}
& \lambda^{|l_2+e_3|}\\
\lambda^{|l_3+e_1|} & \lambda^{|l_3+e_2|}
& \lambda^{|l_3+e_3|}
\end{array}
\right)
\label{Ye}
\eeq

%%%%%%%%%%%%%%%%%%%%%%%%%%%%%%%%%%%%%%%%%%%%%%%%%%%%%%%%%%%%%%%%%
For the quarks we shall assume a common form for the textures 
of $Y_{u}$ and $Y_{d}$

\beq
Y_{u}\sim\left(\begin{array}{ccc}
            \lambda^8 & \lambda^5 & \lambda^3 \\
            \lambda^7 & \lambda^4 & \lambda^2 \\
            \lambda^5 & \lambda^2 & 1
      \end{array}\right), 
Y_{d}\sim\left(\begin{array}{ccc}
            \lambda^4 & \lambda^3 & \lambda^3 \\
           \lambda^3  & \lambda^2 & \lambda^2 \\
           \lambda    &  1        &  1
        \end{array}\right)
\label{quarks}
\eeq
%%%%%%%%%%%%%%%%%%%%%%%%%%%%%%%%%%%%%%%%%%%%%%%%%%%%%%%%%%%%%%%%%%%

The conditions for achieving bi-maximal mixing with
either a conventional neutrino hierarchy,
or an inverted hierarchy 
then translate into conditions on the
choice of $U(1)$ charges for the different fields.
We have already tabulated the simplest charges consistent with the
SRHND conditions giving a hierarchical spectrum and the LMA MSW
solution Eqs.\ref{SRHNDoffdiag}, \ref{LMAMSWoffdiag} \cite{SK2}. 
In Tables 1 and 2, we give the simplest $U(1)$ charges 
consistent with an inverted neutrino mass hierarchy
Eqs.\ref{invertedcondn}, \ref{invertedcondn2}

 The $U(1)$ charges in Tables 1,2 can be arranged into four categories
(referred to as cases I, II, III and IV) according to the perturbation 
terms present in $m_{LL}$  which can be expressed in leading order,
using Eqs.\ref{mRR}, \ref{Ynu} together with Eqs.\ref{N3Dirac},
\ref{MRRoff-diag},\ref{matrix5} as
\beq
m_{LL}\sim 
\left(\begin{array}{ccc}
        \delta & 1 & 1 \\
        1      & \epsilon & \epsilon \\
        1      & \epsilon & \epsilon 
         \end{array}\right)
\eeq
where $\delta=\epsilon=\lambda^{4}$ in cases I,II, $\delta=\lambda^{6}$,
$\epsilon=\lambda^{4}$ in case III, and $\delta=\lambda^{4}$, $\epsilon=
\lambda^{6}$ in case IV. Again one can differentiate cases I and II 
 with respect to the textures of charged lepton Yukawa matrix
as shown  in Table 3. This may lead to 
different solar mixing angles at the end. 

%%%%%%%%%%%%%%%%%%%%%%%%%%%%%%%%%%%%%%%%%%%%%%%%%%%%%%%%%%%%%%%%%%%%%%%%%%%
{\small
\begin{table}[tbp]
\hfil
\begin{tabular}{|l|lllllll|} \hline \hline
     Case  &     $ l_{1}$&$l_{2}$&$l_{3}$&$n_{1}$&$n_{2}$&$n_{3}$&$\sigma$ \\ \hline\hline
      I   & -3 & 3 & 3 & 2 & -3 & 0 & 1  \\
         & -3 & 3 & 3 & 2 & -3 & 1 & -1 \\
         & -3 & 3 & 3 & 3 & -2 & -1 & 1 \\
         & -3 & 3 & 3 & 2 & -2 & 0 & 0 \\
         & -3 & 3 & 3 & 3 & -2 & 0 & -1 \\
         &  3 &-3& -3& -3& 2& 0& 1\\ 
         &  3 &-3& -3& -3& 2& 1& -1 \\
         &  3& -3& -3& -2& 2& 0& 0 \\
         &  3& -3& -3 &-2& 3& -1& 1\\ 
         &  3& -3& -3& -2& 3& 0& -1\\ 
         &  3& -3& -3& -2& 3& 1& -2 \\\hline
     II &       -3& 1& 1& 3& -3& 1& -2 \\
         &  -3& 1& 1& 3& -2& 1& -2 \\
         &  -3 &1& 1& 3& -1& 1& -2 \\
         & -2 & 2 & 2 & 2 & -3 & 0 & 0 \\
         & -2 & 2 & 2 & 2 & -2 & 0 & 0 \\
         & -2 & 2 & 2 & 3 & -2 & 0 & 0 \\
         & -1 & 3 & 3 & 1 & -3 & -1 & 2 \\
         & -1 & 3 & 3 & 2 & -3 & -1 & 2 \\
         & -1 & 3 & 3 & 3 & -3 & -1 & 2 \\
         &  1 & -3&-3 &-3 & 3  &  1 &-2 \\
         &  1 & -3&-3 &-2 & 3  &  1 &-2 \\
         &  1 & -3&-3 &-1 & 3 &   1 & -2 \\
         &  2 & -2& -2&-3 & 2 & 0 & 0  \\
         &  2 & -2& -2& -2& 2 & 0 & 0  \\
         &  2 & -2& -2& -2& 3 & 0 & 0 \\
         &  3 & -1& -3& -3& 2& -1& 2 \\
         &  3 & -1& -1& -3& 1&-1&2 \\
         &  3 & -1& -1& -3& 2&-1&2 \\
         &  3 & -1& -1& -3& 3& -1& 2 \\ \hline\hline
\end{tabular}
\hfil
\caption{\footnotesize Examples of charges (cases I and II) 
which satisfy the 
conditions in Eqs.\ref{invertedcondn},\ref{invertedcondn2}.}
\end{table}
}
%%%%%%%%%%%%%%%%%%%%%%%%%%%%%%%%%%%%%%%%%%%%%%%%%%%%%%%%%%%%
{\small
\begin{table}[tbp]
\hfil
\begin{tabular}{|l|lllllll|} \hline \hline
     Case &     $ l_{1}$&$l_{2}$&$l_{3}$&$n_{1}$&$n_{2}$&$n_{3}$&$\sigma$ \\ \hline\hline
         
     III &     -3 & 3 & 3 & 2 & -3& -1& 2 \\
         & -3 & 3 & 3 & 2 & -3 & 0 & 0 \\
         & -3 & 3 & 3 & 3 & -3 & -1& 1 \\
         & 3  & -3&-3 & -3 & 3 & 1 & -1 \\
         &  3 & -3 & -3& -2& 3& 0&0 \\ \hline
      IV &     -3& 3  &  3 & 3&  -2 & 0 & 0 \\
         &  -3 & 3 & 3 & 3 & -2 & 1 & -2 \\
         &  3 &-3 &-3& -3& 2& -1& 2 \\
         &  3& -3& -3& -3& 2& 0& 0   \\
         &  3& -3& -3& -3& 3& -1& 1 \\ \hline\hline

 \end{tabular}
\hfil
\caption{\footnotesize Examples of charges (cases III and IV) 
which satisfy the conditions 
conditions in Eqs.\ref{invertedcondn},\ref{invertedcondn2}.}
\end{table}
}

%%%%%%%%%%%%%%%%%%%%%%%%%%%%%%%%%%%%%%%%%%%%%%%%%%%%%%%%%%%%%%%%%%%%

\section{Renormalisation Group Analysis}
 We now turn to a renormalisation group study for calculating 
radiative corrections to neutrino masses and mixing angles at 
low energies (see \cite{KS} and references therein). 
We pick up four representative examples, one each 
from each case listed in Tables 1,2. Our detailed
 procedure and methods follow closely that used in the case of SRHND in 
 \cite{KS}, which we summarise briefly. From \cite{KS} we make  
use of the renormalisation
group equations (RGEs) to one-loop order for Yukawa matrices and 
three gauge couplings, in the minimal supersymmetric standard model
with three right-handed neutrinos, including the effects of the heavy
neutrino thresholds. Using the textures
in Tables 3,4, and Eq.\ref{quarks} we first run the quanties
$Y^{f} (f=u,d,e,\nu )$ , $Y_{RR}$ and gauge couplings $g_{1,2,3}$ 
from the GUT scale
$M_{U}=2.0\times10^{16}GeV$ down to the lightest right-handed heavy neutrino
mass scale $M_{R1}$. We take the effects of the other heavy right-handed 
neutrino mass thresholds in successive steps in running the RGEs. Using 
the see-saw formula in the standard way, and inverting $M_{RR}$
numerically, the left-handed Majorana 
neutrino mass matrix $m_{LL}$
is calculated at the scale $M_{R1}$. 
The corresponding lepton mixing matrix 
$V_{MNS}=V_{eL}V_{\nu L}^{\dagger}$ is 
calculated after performing the diagonalisation of 
$m_{LL}^{diag}=V_{\nu L}m_{LL}V_{\nu L}^
{\dagger}$ and $Y_{e}^{diag}=V_{eL}Y_{e}V_{eR}^{\dagger}$ at this scale. From 
$V_{MNS}$ mixing matrix we extract $S_{sol}=\sin^{2}2\theta_{12}$
 as well as $S_{at}=\sin^{2}2\theta_{23}$ as outlined in \cite{KS}. We also 
calculate the left-handed Majorana neutrino mass matrix and the same 
mixing parameters at the GUT scale for 
providing a meaningful comparison of radiative corrections at different 
energy scales  though the see-saw mechanism in principle, does not operate 
at scales above the lightest
right-handed neutrino mass $M_{R1}$. In the next step while moving from 
$M_{R1}$ scale to low energy scale $m_{t}$,  we run the RGEs for the 
coefficient $\kappa$ of the dimension 5 neutrino mass operator 
in the diagonal charged lepton basis \cite{KS}, 
which is interpreted as see-saw 
mass matrix $m_{LL}^{\prime}(M_{R1})=v^{2}_{2}\kappa^{\prime}(M_{R1})$ 
where $\kappa^{\prime}=V_{eL}\kappa V_{eL}^{\dagger}$. This in turn gives
$m_{LL}^{\prime}(m_{t})$,
\beq
m_{LL}^{\prime}(m_{t})=e^{\frac{6}{5}I_{g1}}e^{6I_{g2}}e^{-6I_{t}}
     \left(\begin{array}{ccc}
            m_{LL11}^{\prime}(M_{R1}) & m_{LL12}^{\prime}(M_{R1}) 
& m_{LL13}^{\prime}(M_{R1})e^{-I_{\tau}} \\
            m_{LL21}^{\prime}(M_{R1}) & m_{LL22}^{\prime}(M_{R1}) 
& m_{LL23}^{\prime}(M_{R1})e^{-I_{\tau}} \\
            m_{LL31}^{\prime}(M_{R1})e^{-I_{\tau}} 
                                     & m_{LL32}^{\prime}(M_{R1})e^{-I_{\tau}}
                                    & m_{LL33}^{\prime}(M_{R1})e^{-2I_{\tau}}
          \end{array}\right)
\eeq         

and 
\beq
I_{g_i}=\frac{1}{16\pi^{2}}\int_{\ln m_{t}}^{\ln M_{R1}}g_i^{2}(t)dt,\
\ \ 
I_{f}=\frac{1}{16\pi^{2}}\int_{\ln m_{t}}^{\ln M_{R1}}h_f^{2}(t)dt
\eeq
where $f=t,{\tau} $. 
We also calculate $V_{MNS}(m_{t})=V_{\nu L}^{\prime\dagger}$ 
where $V_{\nu L}^{\prime}$ is the matrix which diagonalises $m_{LL}^{\prime}$.
We then obtain the neutrino masses $m_{LL}^{\prime diag}=diag(m_{\nu1},
m_{\nu2},m_{\nu3})$ and mixing angle parameters  $S_{sol}, S_{at}$ as before.
The parameters $I_{f}$  along with the input textures of 
$Y^{e}, Y^{\nu}, Y_{RR}$
are given in Tables 3,4 which are needed in getting the results in 
Tables 5-8.

%%%%%%%%%%%%%%%%%%%%%%%%%%%%%%%%%%%%%%%%%%%%%%%%%%%%%%%%%%%%%%%%%%%%%%

{\small
\begin{table}[tbp]
\hfil
\begin{tabular}{|l|l|l|} \hline\hline
Parameter & Case I  & Case II   \\ \hline\hline
$U(1)$  & $l_{1,2,3}=-3, 3, 3$ & $l_{1,2,3}=-2, 2, 2$  \\ 
charges & $n_{1,2,3}=2, -3,0$ & $n_{1,2,3}=2,-2,0$   \\
        & $e_{1,2,3}=-5,-1,-3$ & $e_{1,2,3}=-6,0,-2$   \\
        & $\sigma=1$ & $\sigma=0$    \\ \hline
$Y^{e}$ & $\left(\begin{array}{ccc}
                             a_{11}\lambda^8 & a_{12}\lambda^4 & a_{13}\lambda^6 \\
                             a_{21}\lambda^2 & a_{22}\lambda^2 & a_{23}          \\
                             a_{31}\lambda^2 & a_{32}\lambda^2 & a_{33}          
                      \end{array}\right)$
                 & $ \left(\begin{array}{ccc}
                             a_{11}\lambda^8 & a_{12}\lambda^2 & a_{13}\lambda^4  \\
                             a_{21}\lambda^4 & a_{22}\lambda^2 & a_{23}            \\
                             a_{31}\lambda^4 & a_{32}\lambda^2 & a_{33}    
                       \end{array}\right)$
                                  \\ \hline
$a_{ij}$ & $\left(\begin{array}{ccc}
                             1.0 & 1.0 & 0.8 \\
                             1.0 & 2.5  & 0.35 \\
                             0.8 & 1.0 & 2.0 
                       \end{array}\right)$
            & $ \left(\begin{array}{ccc}
                             1.0 & 0.8 & 1.0 \\
                             1.0 & 3.25 & 0.35   \\
                             1.0 & 1.0 & 2.15
                        \end{array}\right)$ 
                              \\ \hline
$Y^{\nu}$ & $ \left(\begin{array}{ccc}
                             \lambda & \lambda^6 & \lambda^3 \\
                             \lambda^5 & 1 & \lambda^3  \\
                             \lambda^5 & 1 & \lambda^3
                      \end{array}\right)$
                 & $ \left(\begin{array}{ccc}
                                 1  & \lambda^4 & \lambda^2  \\
                             \lambda^4& 1 & \lambda^2  \\
                             \lambda^4 &1 & \lambda^2
                       \end{array}\right)$   
                                 \\ \hline
$\frac{M_{RR}}{<\Sigma>}$ &  $ \left(\begin{array}{ccc}
                             \lambda^5 & 1 & \lambda^3 \\
                             1 & \lambda^5 & \lambda^2 \\
                             \lambda^3 & \lambda^2& \lambda
                         \end{array}\right)$
                 & $ \left(\begin{array}{ccc}
                             \lambda^4 & 1 & \lambda^2 \\
                             1 & \lambda^4 & \lambda^2 \\
                             \lambda^2 & \lambda^2 & 1 
                         \end{array}\right)$
                            \\ \hline
$<\Sigma>$      & $0.57\times10^{14}GeV $ & $2.6\times10^{14}GeV$  \\ \hline
$M_{R_{1}}$        & $1.25\times10^{13}GeV$ & $2.6\times10^{14}GeV$  \\ \hline
$I_{g_{1,2}}$ & $0.0433,0.0683$ & $0.0542,0.0809$   \\
$I_{t,\tau}$  & $0.0968,0.0478$ & $0.1107,0.0602$   \\
$I_{\mu,e}$   & $0.0002,1.04\times10^{-8}$ & $0.0004,3.89\times10^{-9}$  \\ \hline\hline

\end{tabular}
\hfil
\caption{\footnotesize For cases I and II: Textures of the Yukawa 
couplings of Dirac neutrino mass, right-handed Majorana neutrino mass, 
and also other relevant parameters needed for the numerical estimation of 
left-handed Majorana neutrino masses at low energies through seesaw 
mechanism. Here $<\Sigma>$ is taken as a free parameter and $M_{R1}$ is 
the lowest threshold scale in $M_{RR}$.
As noted earlier, the right-handed neutrinos may be permuted
according to $1\rightarrow 2,2\rightarrow 3,3\rightarrow 1$
so that the large elements in $Y^{\nu}$ appear in the
23 and 33 positions, which may be more natural from the
viewpoint of unified theories where a large 33 element is expected.
}
\end{table}
}

%%%%%%%%%%%%%%%%%%%%%%%%%%%%%%%%%%%%%%%%%%%%%%%%%%%%%%%%%%%%%%%%%

{\small
\begin{table}[tbp]
\hfil
\begin{tabular}{|l|l|l|} \hline\hline
Parameter & Case III  & Case IV   \\ \hline\hline
$U(1)$  & $l_{1,2,3}=-3,3,3$ & $l_{1,2,3}=3,-3,-3$  \\ 
charges & $n_{1,2,3}=2,-3,-1$ & $n_{1,2,3}=-3,3,-1$   \\
        & $e_{1,2,3}=-5,-1,-3$ & $e_{1,2,3}=5,1,3$   \\
        & $\sigma=2$ & $\sigma=1$    \\ \hline
$Y^{e}$ & $\left(\begin{array}{ccc}
                             a_{11}\lambda^8 & a_{12}\lambda^4 & a_{13}\lambda^6 \\
                             a_{21}\lambda^2 & a_{22}\lambda^2 & a_{23}          \\
                             a_{31}\lambda^2 & a_{32}\lambda^2 & a_{33}          
                      \end{array}\right)$
                 & $ \left(\begin{array}{ccc}
                             a_{11}\lambda^8 & a_{12}\lambda^4 & a_{13}\lambda^6  \\
                             a_{21}\lambda^2 & a_{22}\lambda^2 & a_{23}            \\
                             a_{31}\lambda^2 & a_{32}\lambda^2 & a_{33}    
                       \end{array}\right)$
                                  \\ \hline
$a_{ij}$ & $\left(\begin{array}{ccc}
                             1.0 & 1.0 & 0.8 \\
                             1.0 & 2.5  & 0.35 \\
                             0.8 & 1.0 & 2.0 
                       \end{array}\right)$
            & $ \left(\begin{array}{ccc}
                             1.0 & 1.0 & 0.8 \\
                             1.0 & 2.5 & 0.35   \\
                             0.8 & 1.0 & 2.0
                        \end{array}\right)$ 
                              \\ \hline
$Y^{\nu}$ & $ \left(\begin{array}{ccc}
                             \lambda & \lambda^6 & \lambda^4 \\
                             \lambda^5 & 1 & \lambda^2  \\
                             \lambda^5 & 1 & \lambda^2
                      \end{array}\right)$
                 & $ \left(\begin{array}{ccc}
                                 1  & \lambda^6 & \lambda^2  \\
                             \lambda^6& 1 & \lambda^4  \\
                             \lambda^6 &1 & \lambda^4
                       \end{array}\right)$   
                                 \\ \hline
$\frac{M_{RR}}{<\Sigma>}$ &  $ \left(\begin{array}{ccc}
                             \lambda^6 & \lambda & \lambda^3 \\
                             \lambda & \lambda^4 & \lambda^2 \\
                             \lambda^3 & \lambda^2& 1
                         \end{array}\right)$
                 & $ \left(\begin{array}{ccc}
                             \lambda^5 & \lambda & \lambda^3 \\
                             \lambda & \lambda^7 & \lambda^3 \\
                             \lambda^3 & \lambda^3 & \lambda 
                         \end{array}\right)$
                            \\ \hline
$<\Sigma>$      & $2.592\times10^{14}GeV $ & $11.782\times10^{14}GeV$  \\ \hline
$M_{R_{1}}$        & $5.72\times10^{13}GeV$ & $11.782\times10^{14}GeV$  \\ \hline
$I_{g_{1,2}}$ & $0.0470,0.0728$ & $0.0553,0.0824$   \\
$I_{t,\tau}$  & $0.1016, 0.0518$ & $0.1125,0.0620$   \\
$I_{\mu,e}$   & $0.0002,1.11\times10^{-8}$ & $0.0004,3.64\times10^{-9}$  \\ \hline\hline
\end{tabular}
\hfil
\caption{\footnotesize For cases III and IV: Textures of the Yukawa couplings
of Dirac neutrino mass, right-handed Majorana neutrino mass, and other 
relevant parameters needed for the numerical estimation of left-handed
Majorana neutrino masses at low energies through see-saw mechanism. Here
$<\Sigma>$ is taken as a free parameter and $M_{R1}$ is the lowest 
threshold scale in $M_{RR}$.
As noted earlier, the right-handed neutrinos may be permuted
according to $1\rightarrow 2,2\rightarrow 3,3\rightarrow 1$
so that the large elements in $Y^{\nu}$ appear in the
23 and 33 positions, which may be more natural from the
viewpoint of unified theories where a large 33 element is expected.
}
\end{table}
}

%%%%%%%%%%%%%%%%%%%%%%%%%%%%%%%%%%%%%%%%%%%%%%%%%%%%%%%%%%%%%%%%%%%

{\small
\begin{table}[tbp]
\hfil
\begin{tabular}{|l|l|} \hline
      Scale $ \mu=M_{U}=2.0\times10^{16}GeV$    &  Scale $\mu=M_{U}=2.0\times10^{16}GeV$ \\ \hline
$Y^e = \left(\begin{array}{lll}
                               1.976.10^{-6} & 8.433.10^{-4} & 3.265.10^{-5} \\
                               1.742.10^{-2} & 4.356.10^{-2} & 0.126 \\
                               1.394.10^{-2} & 1.742.10^{-2} & 0.7200
                           \end{array}\right)$
        & $V_{eL}=\left(\begin{array}{lll}
                               -0.999 & 0.018 & -0.003 \\
                               -0.019 & -0.984 & 0.175 \\
                               0.000 & 0.175 & 0.985 
                            \end{array}\right)$
                               \\
$Y^{e}_{diag}=diag(2.9097.10^{-4}, 4.252.10^{-2}, 0.73155)$ & $m_{e}/m_{\mu},m_{\mu}/m_{\tau}=0.0068, 0.0581$
                                      
                                \\ 
$m_{LL}^{\nu}=$  & $V_{MNS}=V_{eL}V_{{\nu}L}^{\dagger}=$     \\
$ \left(\begin{array}{lll}
                               1.511.10^{-6} &  1.332.10^{-2} &  1.332.10^{-2} \\
                               1.332.10^{-2} &  3.121.10^{-5} & 3.121.10^{-5} \\
                               1.332.10^{-2} &  3.121.10^{-5} & 3.121.10^{-5}
                             \end{array}\right)$
        & $ \left(\begin{array}{lll}
                              -0.699 &-0.715 & -0.015 \\
                               0.418 & 0.391 &  0.820 \\
                               0.580 &-0.579 &  0.573
                              \end{array}\right)$

                                    \\
$m_{LL}^{diag}=diag(0.085782, -0.085492,-3.17.10^{-17})$ & $ S_{sol}=0.9995,S_{at}=0.8818$
                                    \\ \hline
  Scale $\mu=M_{R_{1}}=1.25\times10^{13}GeV$ &  Scale $\mu=M_{R_{1}}=1.25\times10^{13}GeV $ 
                                    \\ \hline
$Y^e=\left(\begin{array}{lll}
                               1.561.10^{-6} & 7.995.10^{-4} & 2.292.10^{-5}  \\
                               1.591.10^{-2} & 4.020.10^{-2} & 9.999.10^{-2}  \\
                               1.181.10^{-2} & 1.424.10^{-2} & 0.629  
                            \end{array}\right)$
         & $V_{eL}=\left(\begin{array}{lll}
                               -0.999 & 0.019 & -0.003 \\
                               -0.019 & -0.987 & 0.159 \\
                               0.000  & 0.159 & 0.987 
                             \end{array}\right)$
                                   \\
$Y^{e}_{diag}=diag(2.7605.10^{-4}, 3.9926.10^{-2},0.63712)$ & $m_{e}/m_{\mu},m_{\mu}/m_{\tau}=0.0069,0.0627$
                                        
                                    \\
$m_{LL}^{\nu}=$  & $ V_{MNS}=V_{eL}V_{{\nu}L}^{\dagger}=$  \\ 

 $\left(\begin{array}{lll}
                               5.755.10^{-6} & 5.889.10^{-2} & 5.792.10^{-2} \\
                               5.889.10^{-2} & 1.004.10^{-4} & 9.885.10^{-5} \\
                               5.792.10^{-2} & 9.885.10^{-5} & 9.731.10^{-5}
                              \end{array}\right)$
        & $\left(\begin{array}{lll}
                              -0.698 & -0.715 & -0.015 \\
                              -0.432 &  0.405 & 0.806 \\
                               0.570 & -0.569 &  0.592
                              \end{array}\right)$
                                   \\ 
 $m_{LL}^{diag}=diag(0.082704,-0.082500,9.7.10^{-10})$ & $S_{sol}=0.9994,S_{at}=0.9107$ \\ \hline
Scale $\mu= m_{t}=175 GeV$ &  Scale $\mu=m_{t}=175 GeV$   \\ \hline

$Y^{e}_{diag}=diag(2.391.10^{-4}, 3.456.10^{-2}, 0.4774)$ & $m_{e}/m_{\mu},m_{\mu}/m_{\tau}=0.0069,0.0724$
                                       \\
 $m_{LL}^{\prime\nu}=$  & $ V_{MNS}=V_{{\nu}L}^{\dagger}=$   \\
$\left(\begin{array}{lll}
                               -1.633.10^{-3} & 4.340.10^{-2} & -5.634.10^{-2} \\
                               4.340.10^{-2} & 1.691.10^{-3}  & -1.140.10^{-3} \\
                              -5.634.10^{-2} & -1.140.10^{-3} &  1.119.10^{-4} 
                              \end{array}\right)$
         & $\left(\begin{array}{lll}
                               -0.698 & -0.716 & -0.015 \\
                              -0.445 & 0.418 &  0.792 \\
                               0.560 & -0.560 & 0.610 
                              \end{array}\right)$
                                 \\
 $m_{LL}^{\prime diag}=diag(0.071229,-0.071059,-1.6.10^{-9})$ & $S_{sol}=0.9994,S_{at}=0.9352$ \\ \hline

\end{tabular}
\hfil
\caption{\footnotesize Case I: Left-handed Majorana neutrino mass matrix
and mixing matrix at different energy scales $M_{U}$, $M_{R1}$ and $m_{t}$.
 Neutrino masses are expressed in eV.
Note that we take the convention $|m_{\nu_1}|>|m_{\nu_2}|$,
which fixes the ordering of the 1st two columns of $V_{MNS}$.
}
\end{table}
}
%%%%%%%%%%%%%%%%%%%%%%%%%%%%%%%%%%%%%%%%%%%%%%%%%%%%%%%%%%%%

{\small
\begin{table}[tbp]
\hfil
\begin{tabular}{|l|l|} \hline
      Scale $ \mu=M_{U}=2.0\times10^{16}GeV$    &  Scale $\mu=M_{U}=2.0\times10^{16}GeV$ \\ \hline
$Y^e = \left(\begin{array}{lll}
                               1.838.10^{-6} & 1.297.10^{-2} & 7.845.10^{-4} \\
                               7.845.10^{-4} & 5.268.10^{-2} & 0.1172 \\
                               7.845.10^{-4} & 1.621.10^{-2} & 0.7200
                           \end{array}\right)$
        & $V_{eL}=\left(\begin{array}{lll}
                               -0.967 & 0.250 & -0.040 \\
                               -0.253 & -0.954 & 0.158 \\
                               0.002 & 0.163 & 0.987 
                            \end{array}\right)$
                               \\
$Y^{e}_{diag}=diag(1.634.10^{-4}, 5.103.10^{-2}, 0.7299)$ & $m_{e}/m_{\mu},m_{\mu}/m_{\tau}=0.0032, 0.0699$
                                      
                                \\ 
$m_{LL}^{\nu}=$  & $V_{MNS}=V_{eL}V_{{\nu}L}^{\dagger}=$     \\
$ \left(\begin{array}{lll}
                               1.419.10^{-4} &  6.055.10^{-2} &  6.055.10^{-2} \\
                               6.055.10^{-2} &  1.419.10^{-4} & 1.419.10^{-4} \\
                               6.055.10^{-2} &  1.419.10^{-4} & 1.419.10^{-4}
                             \end{array}\right)$
        & $ \left(\begin{array}{lll}
                               0.578 & 0.790 & -0.205 \\
                               0.577 &-0.219 & 0.787 \\
                              -0.576 & 0.573 & 0.582 
                              \end{array}\right)$

                                    \\
$m_{LL}^{diag}=diag(0.085846, -0.085424,1.3.10^{-17})$ & $ S_{sol}=0.9091,S_{at}=0.9149$
                                    \\ \hline
  Scale $\mu=M_{R_{1}}=2.6\times10^{14}GeV$ &  Scale $\mu=M_{R_{1}}=2.6\times10^{14}GeV $ 
                                    \\ \hline
$Y^e=\left(\begin{array}{lll}
                               1.623.10^{-6} & 1.254.10^{-2} & 6.775.10^{-4}  \\
                               7.484.10^{-4} & 5.072.10^{-2} & 1.040.10^{-1}  \\
                               7.281.10^{-4} & 1.457.10^{-2} & 0.676  
                            \end{array}\right)$
         & $V_{eL}=\left(\begin{array}{lll}
                               -0.968 & 0.250 & -0.037 \\
                               -0.253 & -0.956 & 0.150 \\
                               0.002  & 0.154 & 0.988 
                             \end{array}\right)$
                                   \\
$Y^{e}_{diag}=diag(1.58.10^{-4}, 4.950.10^{-2},0.6843)$ & $m_{e}/m_{\mu},m_{\mu}/m_{\tau}=0.003195,0.0723$
                                        
                                    \\
$m_{LL}^{\nu}=$  & $ V_{MNS}=V_{eL}V_{{\nu}L}^{\dagger}=$  \\ 

 $\left(\begin{array}{lll}
                               7.659.10^{-5} & 5.905.10^{-2} & 5.851.10^{-2} \\
                               5.905.10^{-2} & 9.409.10^{-5} & 9.327.10^{-5} \\
                               5.851.10^{-2} & 9.327.10^{-5} & 9.246.10^{-5}
                              \end{array}\right)$
        & $\left(\begin{array}{lll}
                              -0.577 & -0.791 & -0.203 \\
                              -0.584 &  0.227 & 0.779  \\
                               0.571 & -0.568 &  0.593
                              \end{array}\right)$
                                   \\ 
 $m_{LL}^{diag}=diag(0.083260,-0.082997,2.1.10^{-11})$ & $S_{sol}=0.9067,S_{at}=0.9291$ \\ \hline
Scale $\mu= m_{t}=175 GeV$ &  Scale $\mu=m_{t}=175 GeV$   \\ \hline

$Y^{e}_{diag}=diag(1.332.10^{-4}, 4.16.10^{-2}, 0.4798)$ & $m_{e}/m_{\mu},m_{\mu}/m_{\tau}=0.0032,0.0868$
                                       \\
 $m_{LL}^{\prime\nu}=$  & $ V_{MNS}=V_{{\nu}L}^{\dagger}=$   \\
$\left(\begin{array}{lll}
                               -2.164.10^{-2} & 3.837.10^{-2} & -5.432.10^{-2} \\
                               3.837.10^{-2} & 2.158.10^{-2}  & -1.433.10^{-2} \\
                              -5.432.10^{-2} & -1.433.10^{-2} & 2.628.10^{-4} 
                              \end{array}\right)$
         & $\left(\begin{array}{lll}
                               -0.573 & -0.795 & -0.198 \\
                              -0.601 & 0.243 &  0.762 \\
                               0.558 & -0.555 & 0.617 
                              \end{array}\right)$
                                 \\
 $m_{LL}^{\prime diag}=diag(0.071479,-0.071278,4.5.10^{-8})$ & $S_{sol}=0.8996,S_{at}=0.9568$ \\ \hline

\end{tabular}
\hfil
\caption{\footnotesize Case II: Left-handed Majorana neutrino mass 
matrix and mixing matrix at different energy scales $M_{U}$, $M_{R1}$ 
and $m_{t}$. Neutrino masses are expressed in eV.
Note that we take the convention $|m_{\nu_1}|>|m_{\nu_2}|$,
which fixes the ordering of the 1st two columns of $V_{MNS}$.
}
\end{table}
}

%%%%%%%%%%%%%%%%%%%%%%%%%%%%%%%%%%%%%%%%%%%%%%%%%%%%%%%%%%%%%%%%%%%%%%%%

{\small
\begin{table}[tbp]
\hfil
\begin{tabular}{|l|l|} \hline
      Scale $ \mu=M_{U}=2.0\times10^{16}GeV$    &  Scale $\mu=M_{U}=2.0\times10^{16}GeV$ \\ \hline
$Y^e = \left(\begin{array}{lll}
                               1.976.10^{-6} & 8.433.10^{-4} & 3.265.10^{-5} \\
                               1.742.10^{-2} & 4.356.10^{-2} & 0.126 \\
                               1.394.10^{-2} & 1.742.10^{-2} & 0.7200
                           \end{array}\right)$
        & $V_{eL}=\left(\begin{array}{lll}
                               -0.999 & 0.018 & -0.003 \\
                               -0.019 & -0.984 & 0.175 \\
                               0.000 & 0.175 & 0.985 
                            \end{array}\right)$
                               \\
$Y^{e}_{diag}=diag(2.910.10^{-4}, 4.252.10^{-2}, 0.7316)$ & $m_{e}/m_{\mu},m_{\mu}/m_{\tau}=0.0068, 0.0581$
                                      
                                \\ 
$m_{LL}^{\nu}=$  & $V_{MNS}=V_{eL}V_{{\nu}L}^{\dagger}=$     \\
$ \left(\begin{array}{lll}
                               3.323.10^{-7} &  6.055.10^{-2} &  6.055.10^{-2} \\
                               6.055.10^{-2} &  1.419.10^{-4} & 1.419.10^{-4} \\
                               6.055.10^{-2} &  1.419.10^{-4} & 1.419.10^{-4}
                             \end{array}\right)$
        & $ \left(\begin{array}{lll}
                              -0.699 &-0.715 & -0.015 \\
                              -0.418 & 0.391 &  0.820 \\
                               0.580 &-0.579 &  0.573
                              \end{array}\right)$

                                    \\
$m_{LL}^{diag}=diag(0.085779, -0.085495,-1.08.10^{-19})$ & $ S_{sol}=0.9995,S_{at}=0.8818$
                                    \\ \hline
  Scale $\mu=M_{R_{1}}=5.72\times10^{13}GeV$ &  Scale $\mu=M_{R_{1}}=5.72\times10^{13}GeV $ 
                                    \\ \hline
$Y^e=\left(\begin{array}{lll}
                               1.626.10^{-6} & 8.069.10^{-4} & 2.438.10^{-5}  \\
                               1.607.10^{-2} & 4.058.10^{-2} & 1.017.10^{-1}  \\
                               1.207.10^{-2} & 1.458.10^{-2} & 0.642  
                            \end{array}\right)$
         & $V_{eL}=\left(\begin{array}{lll}
                               -0.999 & 0.019 & -0.003 \\
                               -0.019 & -0.987 & 0.159 \\
                               0.000  & 0.159 & 0.987 
                             \end{array}\right)$
                                   \\
$Y^{e}_{diag}=diag(2.7856.10^{-4}, 4.0277.10^{-2},0.6502)$ & $m_{e}/m_{\mu},m_{\mu}/m_{\tau}=0.0069,0.0619$
                                        
                                    \\
$m_{LL}^{\nu}=$  & $ V_{MNS}=V_{eL}V_{{\nu}L}^{\dagger}=$  \\ 

 $\left(\begin{array}{lll}
                               8.794.10^{-6} & 5.674.10^{-2} & 5.580.10^{-2} \\
                               5.674.10^{-2} & 1.067.10^{-4} & 1.049.10^{-4} \\
                               5.580.10^{-2} & 1.049.10^{-4} & 1.031.10^{-4}
                              \end{array}\right)$
        & $\left(\begin{array}{lll}
                              -0.699 & -0.715 & -0.015 \\
                              -0.432 &  0.405 & 0.805 \\
                               0.570 & -0.569 &  0.593
                              \end{array}\right)$
                                   \\ 
 $m_{LL}^{diag}=diag(0.079689,-0.07947,3.27.10^{-10})$ & $S_{sol}=0.9994,S_{at}=0.9114$ \\ \hline
Scale $\mu= m_{t}=175 GeV$ &  Scale $\mu=m_{t}=175 GeV$   \\ \hline

$Y^{e}_{diag}=diag(2.39.10^{-4}, 3.455.10^{-2}, 0.4771)$ & $m_{e}/m_{\mu},m_{\mu}/m_{\tau}=0.0069,0.0724$
                                       \\
 $m_{LL}^{\prime\nu}=$  & $ V_{MNS}=V_{{\nu}L}^{\dagger}=$   \\
$\left(\begin{array}{lll}
                               -1.576.10^{-3} & 4.193.10^{-2} & -5.419.10^{-2} \\
                               4.193.10^{-2} & 1.641.10^{-3}  & -1.105.10^{-3} \\
                              -5.419.10^{-2} & -1.105.10^{-3} &  1.173.10^{-4} 
                              \end{array}\right)$
         & $\left(\begin{array}{lll}
                               -0.698 & -0.716 & -0.015 \\
                              -0.446& -0.419 &  0.791 \\
                               -0.560 & -0.559 & 0.612 
                              \end{array}\right)$
                                 \\
 $m_{LL}^{\prime diag}=diag(0.068639,-0.068457, 4.97.10^{-9})$ & $S_{sol}=0.9994,S_{at}=0.9374$ \\ \hline

\end{tabular}
\hfil
\caption{\footnotesize Case III: Left-handed Majorana neutrino mass 
matrix and mixing matrix at different energy scales $M_{U}$, $M_{R1}$ and 
$m_{t}$. Neutrino masses are expressed in eV.
Note that we take the convention $|m_{\nu_1}|>|m_{\nu_2}|$,
which fixes the ordering of the 1st two columns of $V_{MNS}$.
}
\end{table}
}

%%%%%%%%%%%%%%%%%%%%%%%%%%%%%%%%%%%%%%%%%%%%%%%%%%%%%%%%%%%%%%%%%%%%

{\small
\begin{table}[tbp]
\hfil
\begin{tabular}{|l|l|} \hline
      Scale $ \mu=M_{U}=2.0\times10^{16}GeV$    &  Scale $\mu=M_{U}=2.0\times10^{16}GeV$ \\ \hline
$Y^e = \left(\begin{array}{lll}
                               1.976.10^{-6} & 8.433.10^{-4} & 3.265.10^{-5} \\
                               1.742.10^{-2} & 4.356.10^{-2} & 0.126 \\
                               1.394.10^{-2} & 1.742.10^{-2} & 0.7200
                           \end{array}\right)$
        & $V_{eL}=\left(\begin{array}{lll}
                               -0.999 & 0.018 & -0.003 \\
                               -0.019 & -0.984 & 0.175 \\
                               0.000 & 0.175 & 0.985 
                            \end{array}\right)$
                               \\
$Y^{e}_{diag}=diag(2.9097.10^{-4}, 4.252.10^{-2}, 0.7316)$ & $m_{e}/m_{\mu},m_{\mu}/m_{\tau}=0.0068, 0.0581$
                                      
                                \\ 
$m_{LL}^{\nu}=$  & $V_{MNS}=V_{eL}V_{{\nu}L}^{\dagger}=$     \\
$ \left(\begin{array}{lll}
                               1.223.10^{-4} &  6.068.10^{-2} &  6.068.10^{-2} \\
                               6.068.10^{-2} &  3.323.10^{-7} & 3.323.10^{-7} \\
                               6.068.10^{-2} &  3.323.10^{-7} & 3.323.10^{-7}
                             \end{array}\right)$
        & $ \left(\begin{array}{lll}
                              0.699 & 0.714 & -0.015 \\
                               0.418 & -0.392 &  0.820 \\
                               -0.580 & 0.580 &  0.573
                              \end{array}\right)$

                                    \\
$m_{LL}^{diag}=diag(0.0858705, -0.0857475,5.53.10^{-17})$ & $ S_{sol}=0.9996,S_{at}=0.8818$
                                    \\ \hline
  Scale $\mu=M_{R_{1}}=11.782\times10^{14}GeV$ &  Scale $\mu=M_{R_{1}}=11.782\times10^{14}GeV $ 
                                    \\ \hline
$Y^e=\left(\begin{array}{lll}
                               1.768.10^{-6} & 8.170.10^{-4} & 2.780.10^{-5}  \\
                               1.670.10^{-2} & 4.197.10^{-2} & 1.130.10^{-1}  \\
                               1.294.10^{-2} & 1.590.10^{-2} & 0.678  
                            \end{array}\right)$
         & $V_{eL}=\left(\begin{array}{lll}
                               -0.999 & 0.018 & -0.003 \\
                               -0.019 & -0.986 & 0.167 \\
                               0.000  & 0.167 & 0.986
                             \end{array}\right)$
                                   \\
$Y^{e}_{diag}=diag(2.8196.10^{-4}, 4.133.10^{-2},0.688)$ & $m_{e}/m_{\mu},m_{\mu}/m_{\tau}=0.0068,0.0600$
                                        
                                    \\
$m_{LL}^{\nu}=$  & $ V_{MNS}=V_{eL}V_{{\nu}L}^{\dagger}=$  \\ 

 $\left(\begin{array}{lll}
                               4.520.10^{-6} & 5.644.10^{-2} & 5.596.10^{-2} \\
                               5.644.10^{-2} &-2.070.10^{-7} &-2.011.10^{-7} \\
                               5.596.10^{-2} &-2.011.10^{-7} &-1.953.10^{-7}
                              \end{array}\right)$
        & $\left(\begin{array}{lll}
                              -0.699 &  0.715 & 0.015 \\
                              -0.425 & -0.399 & -0.812 \\
                               0.575 &  0.575 &  -0.583
                              \end{array}\right)$
                                   \\ 
 $m_{LL}^{diag}=diag(0.0794824,-0.0794783,-4.0.10^{-13})$ & $S_{sol}=0.9995,S_{at}=0.8975$ \\ \hline
Scale $\mu= m_{t}=175 GeV$ &  Scale $\mu=m_{t}=175 GeV$   \\ \hline

$Y^{e}_{diag}=diag(2.391.10^{-4}, 3.455.10^{-2}, 0.47714)$ & $m_{e}/m_{\mu},m_{\mu}/m_{\tau}=0.0069,0.0724$
                                       \\
 $m_{LL}^{\prime\nu}=$  & $ V_{MNS}=V_{{\nu}L}^{\dagger}=$   \\
$\left(\begin{array}{lll}
                               -1.537.10^{-3} & 4.134.10^{-2} & -5.420.10^{-2} \\
                               4.134.10^{-2} & 1.532.10^{-3}  & -1.007.10^{-3} \\
                              -5.420.10^{-2} & -1.007.10^{-3} &  7.900.10^{-6} 
                              \end{array}\right)$
         & $\left(\begin{array}{lll}
                               -0.699 & 0.715 & -0.015 \\
                              -0.442 & -0.416 &  0.795 \\
                               0.562 & 0.562 & 0.607
                              \end{array}\right)$
                                 \\
 $m_{LL}^{\prime diag}=diag(0.0681955,-0.0681929, 
-3.1.10^{-8})$ & $S_{sol}=0.9995,S_{at}=0.9305$ \\ \hline

\end{tabular}

\hfil
\caption{\footnotesize Case IV: Left-handed Majorana neutrino mass matrix
 and mixing matrix at different energy scales $M_{U}$, $M_{R1}$ and $m_{t}$.
Neutrino masses are expressed in eV.
Note that we take the convention $|m_{\nu_1}|>|m_{\nu_2}|$,
which fixes the ordering of the 1st two columns of $V_{MNS}$.
}
\end{table}
}

%%%%%%%%%%%%%%%%%%%%%%%%%%%%%%%%%%%%%%%%%%%%%%%%%%%%%%%%%%%%%%%%%%%%%%%%%%

The numerical results  are presented in Tables 5-8 for cases I-IV where 
the results in Table 8 is meant for relative comparison only. We define
the measure of the splitting of neutrino masses by  
 $\xi=(m_{\nu1}-m_{\nu2})/m_{\nu2}\approx \frac{1}{2}\frac{\bigtriangleup
m_{12}^{2}}{\bigtriangleup m_{23}^{2}}$ which is found to be 
 decreasing from high energy scale $M_{U}$ to low energy scale $m_{t}$ 
by about $(30,43,20)\%$ for cases I,II and III respectively. 
But the decrease in moving from the scale $M_{U}$
to scale $M_{R1}$ is about $(22,36,17)\%$ which is a significant 
effect.  We get at scale $m_{t}$ the values of 
the measure of the splitting, $\xi=(0.00238,0.00281,0.00266,0.00004)$
which corresponds to 
$m_{\nu_1}-m_{\nu_2}=\xi m_{\nu_2}= (1.7,2.0,1.8,0.03)\times 10^{-4}$ eV
for cases I-IV, which are somewhat on the lower end but within the 
observational range except for case IV (which is completely ruled out).
The low energy absolute values of neutrino masses are also estimated as 
$m_{\nu2}=(0.0711,0.0713,0.0685,0.0682)eV$  and 
$m_{\nu3}\approx 0$ for cases I-IV respectively. There is almost 
a smooth decrease in $m_{\nu2,3}$ by about $20\%$ while moving from 
 scale $M_{U}$ down 
to low energy scale $m_{t}$. There is a mild decrease  
of the solar mixing 
angle parameter $S_{sol}$ while moving from high  scale $M_{U}$ to $m_{t}$
scale  in all cases,  
and its low energy values are $S_{sol}=0.9994,0.8996,0.9994,0.9995$ for 
cases I-IV respectively. These values execpt for the case II, 
are on the higher side which lie outside the allowed range. This depends
 on the texture of charged lepton Yukawa matrix given in Tables 3 and 4.
However the atmospheric mixing angle parameter $S_{at}$ is found to increase
by about $6\%$ from high scale $M_{U}$ to low scale $m_{t}$, of which $2\%$
is in the energy range $M_{U}$ to $M_{R1}$. The low energy values are 
predicted as $S_{at}=0.935,0.957,0.937,0.931$ for cases I-IV respectively.
These values are above the experimental lower bound $S_{at}>0.88$. In our 
numerical analysis the   charged lepton textures also predict  almost 
consistent  hierarchical ratios of charged lepton masses at low scale 
as shown in Tables 5-8. 

\section{Conclusion}
In conclusion, we have
studied models of neutrino masses which naturally give rise to
an inverted mass hierarchy and bi-maximal mixing. The models are 
based on the see-saw
mechanism with three right-handed neutrinos, which 
generates a single mass term
of the form $\nu_e(\nu_{\mu}+\nu_{\tau})$ corresponding
to two degenerate neutrinos $\nu_e$ and $\nu_{\mu}+\nu_{\tau}$,
and one massless neutrino $\nu_{\mu}-\nu_{\tau}$.
Atmospheric neutrino oscillations are accounted for if the
degenerate mass term is about $5\times 10^{-2}$ eV.
Solar neutrino oscillations of the Large Mixing Angle MSW type
arise when small perturbations are included leading
to a mass splitting between the degenerate pair of 
about $(1.7-2.0)\times 10^{-4}$ eV.
We have studied the conditions that such models must satisfy in
the framework of a $U(1)$ family symmetry broken by vector singlets,
and catalogue the simplest examples. 
We distinguished four types of cases, and
then performed a
renormalisation group analysis of the neutrino masses mixing
angles, assuming the supersymmetric standard model and large $\tan
\beta$, for one example from each case.
Cases I,III,IV predict almost maximal solar mixing, and an atmospheric
mixing angle which is near maximal, increasing by 6\% due to RG
running. However case IV predicts a splitting parameter
$\xi$ which is outside the allowed range, therefore this
texture is not favoured for the LMA MSW solution.
Case II gives a somewhat smaller solar mixing angle,
which decreases by about 3\% due to RG running, while the
atmospheric angle increases by about 5\%.
Although these examples predict a large solar mixing angle
$\sin^22\theta_{12}>0.9$, this prediction depends to some extent
on the texture assumed for $Y_e$, since in the absence of 
charged lepton mixing angles, the 12 neutrino mixing is 
almost exactly maximal.
Clearly all cases are stable under radiative corrections,
leading to a natural explanation of bi-maximal mixing
in terms of an experimentally testable inverted hierarchical spectrum.

\end{document}